\documentclass{aa}
\usepackage{graphicx}
\begin{document}
   \title{First results of IBIS/ISGRI Cygnus X-3 monitoring during INTEGRAL PV phase\thanks{Based on observations with INTEGRAL, an ESA project with instruments and science data center funded by ESA member
states (especially the PI countries: Denmark, France, Germany, Italy, Switzerland, Spain), Czech Republic and Poland, and
with the participation of Russia and the USA. }}

   \author{P. Goldoni
          \inst{1}
          \and
          J. M.Bonnet-Bidaud \inst{1}
         \and
        M. Falanga\inst{1} 
          \and
        A. Goldwurm\inst{1}
          }
   \offprints{P. Goldoni}
   \institute{CEA Saclay, DSM/DAPNIA/Service d'Astrophysique, F91191  Gif sur Yvette France \\
              \email{pgoldoni@cea.fr}
             }

   \date{10/09/2003}

   \abstract{ 
We report on preliminary results of IBIS/ISGRI serendipitous observations of Cygnus X-3 in the 15-100 keV energy range during the INTEGRAL Performance and Verification phase. This peculiar microquasar was inside IBIS/ISGRI field of view at a $\sim$ 9$^{\circ}$ distance from the pointing direction during Cygnus X-1 staring observations in November and December 2002. We analyzed observations from 27 November 2002 to 8 December 2002 with an effective on source exposure time of $\sim$ 300 kiloseconds. Cyg X-3 was always significantly detected in the 15-40 and 40-100 keV energy bands during single exposures lasting between 30 minutes and one hour. The source light curve shows the characteristic 4.8-hour modulation with a shape consistent with a standard template. The two light curves' phase zero have no measurable offset and their values are consistent with historical ephemeris. These results show that even at this early stage of the mission, IBIS/ISGRI is capable of producing high quality scientific results on highly off axis, relatively bright targets.
   }

   \maketitle
%

\section{Introduction}

Cygnus X-3 is an enigmatic X-ray binary which does not fit well into any of
the established classes of X-ray Binaries (see Bonnet-Bidaud \& Chardin 1988
for a review). One of its main characteristics
is a 4.8-hour modulation visible in hard X-rays (\cite{hermsen},
\cite{robinson}), soft X-rays (\cite{parsignault}) and infrared (\cite{becklin}, \cite{mason}). If interpreted as the orbital period,
this modulation would imply that Cyg X-3 is a low mass X-ray Binary,
but infrared observations suggest that the donor star is a Wolf-Rayet star
(\cite{vankerkw92}). Due to its heavy absorption, no optical
counterpart has been found. The true nature of the compact object
has not been revealed despite orbital-phase-resolved spectroscopy
performed in infrared (\cite{hanson00}).

\noindent Cyg X-3 is also a strong source of radio emission with different behaviors:
1) quiescence (60-100 mJy), 2) major flaring (greater than 1 Jy)
with quenching (very low fluxes $\sim$ 10-20 mJy) 3) minor flaring (less than 1 Jy)
with partial quenching. During major radio outbursts jet-like
structures have been detected moving at a velo-city of either 0.8 or 0.5 c
depending on whether the jets are one-sided or double-sided (\cite{mioduszewski}, \cite{marti}). This is particularly interesting in the
context of models where hard X-ray emission is originated in a low-level
jet (see e.g. Markoff et al. 2003).

\noindent Soft X-ray emission has been observed to undergo high and low states
in which the various spectral components change with no apparent correlation
with other properties (\cite{white95}). Hard X-ray emission
has been extensively monitored with BATSE in the 20-100 keV band
with the Earth occultation method.
The results have been reported by McCollough et al. (1999) for the
period 1991-1996. This monitoring allowed the construction of light
curves with a 3-day time-scale. The two most distinctive features
of these light curves are extended periods of high flux (150-300 mCrab)
and low flux periods when the source was undetectable. The 20-100
keV flux is correlated with radio emission during flaring activity and
anti-correlated during the quenched radio state. The relationship between
hard X-rays and radio emission may be an indication of the
non thermal nature of the hard X-ray emission. However, BATSE results were
simply fitted with a power law with photon index $\alpha$=3 with no further spectral analysis.

\noindent A unified spectral X-ray model based on GINGA data (2-37 keV)
was proposed by Nakamura et al. (1993). It includes several different
components including a blackbody, a power law with cutoff, an iron
line and dust absorption. This model properly fits soft X-ray data,
but it cannot be used to adress the issue of the origin of X-ray emission
over 40 keV. Indeed, BeppoSAX observations in 1999 indicate deviations
from this behavior (\cite{palazzi}) at hard X-ray energies.
We report here on the first results of INTEGRAL/ISGRI observation of
Cyg X-3 during the Performance and Verification phase. The source
was serendipitously observed for more than one month during PV
phase observations dedicated to Cyg X-1. The combination of high energy
sensitivity and exposure of these observations is unprecedented for Cyg X-3.


\section{Observations}

\noindent The INTEGRAL satellite (Winkler et al. 2003) is an ESA observatory
dedicated to 15 keV-10 MeV $\gamma$-ray observations with concurrent source monitoring
in X-rays (3-35 keV) and in the optical range (V, 500-600 nm). The INTEGRAL payload
consists of two main $\gamma$-ray instruments, the spectrometer SPI 
(Vedrenne et al., 2003) and the imager IBIS,
and of two monitor instruments, the X-ray monitor JEM-X (Lund et al. 2003) and the Optical Monitoring Camera
OMC (Mas-Hesse et al 2003).

\noindent The imaging performances of IBIS (Ubertini et al., 2003) are
characterized by the coupling of its source discrimination capability with a
very wide field of view (FOV), namely 9$^\circ \times  9^\circ$ fully coded,
$29^\circ \times  29^\circ$ partially coded FOV. It consists of two
detection layers, ISGRI and PICsIT, optimized for lower and higher energy sensitivity.
The upper one, ISGRI (Lebrun et al., 2003) is sensitive between
15 keV and 1 MeV while its peak sensitivity is between 15 keV and 200 keV.
The lower one, PICsIT (Di Cocco et al., 2003) is sensitive between
$\sim$ 200 keV and $\sim$ 8 MeV.

\noindent During its Performance and Verification Phase INTEGRAL observed
the Cygnus region from 15 November 2002 to 23 December 2002. The main target
of the PV phase was the Black Hole candidate Cygnus X-1 which was observed
on axis and at different off axis distances (Bazzano et al.,
2003, Laurent et al., 2003). During the main part of these observations
Cyg X-3 was contained in the field of view of the main instruments IBIS and
SPI. We report here on results of observations of the ISGRI camera
of the IBIS telescope.

  \begin{figure}
   \centering
   \includegraphics[width=8 cm]{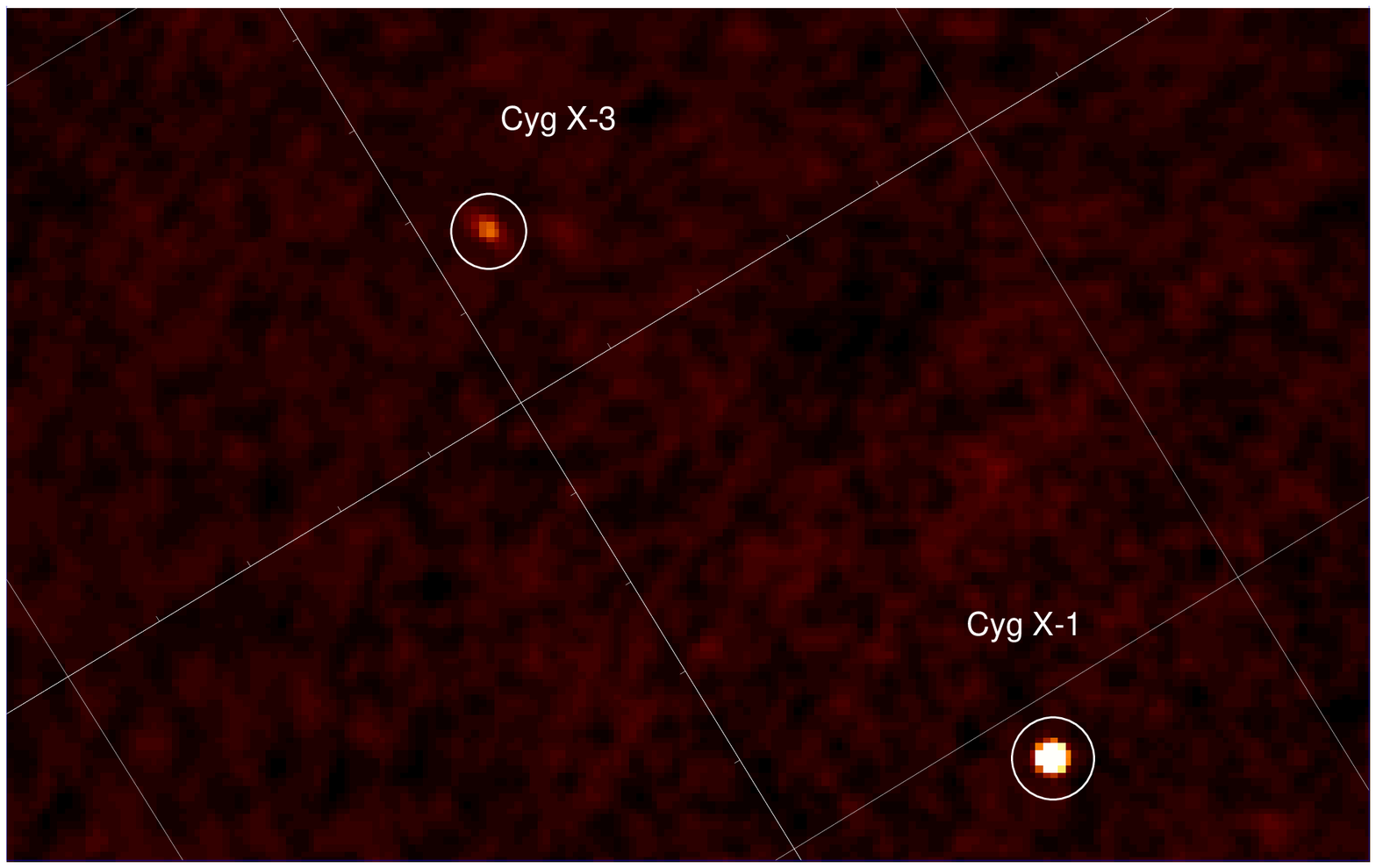}
 \includegraphics[width=8 cm]{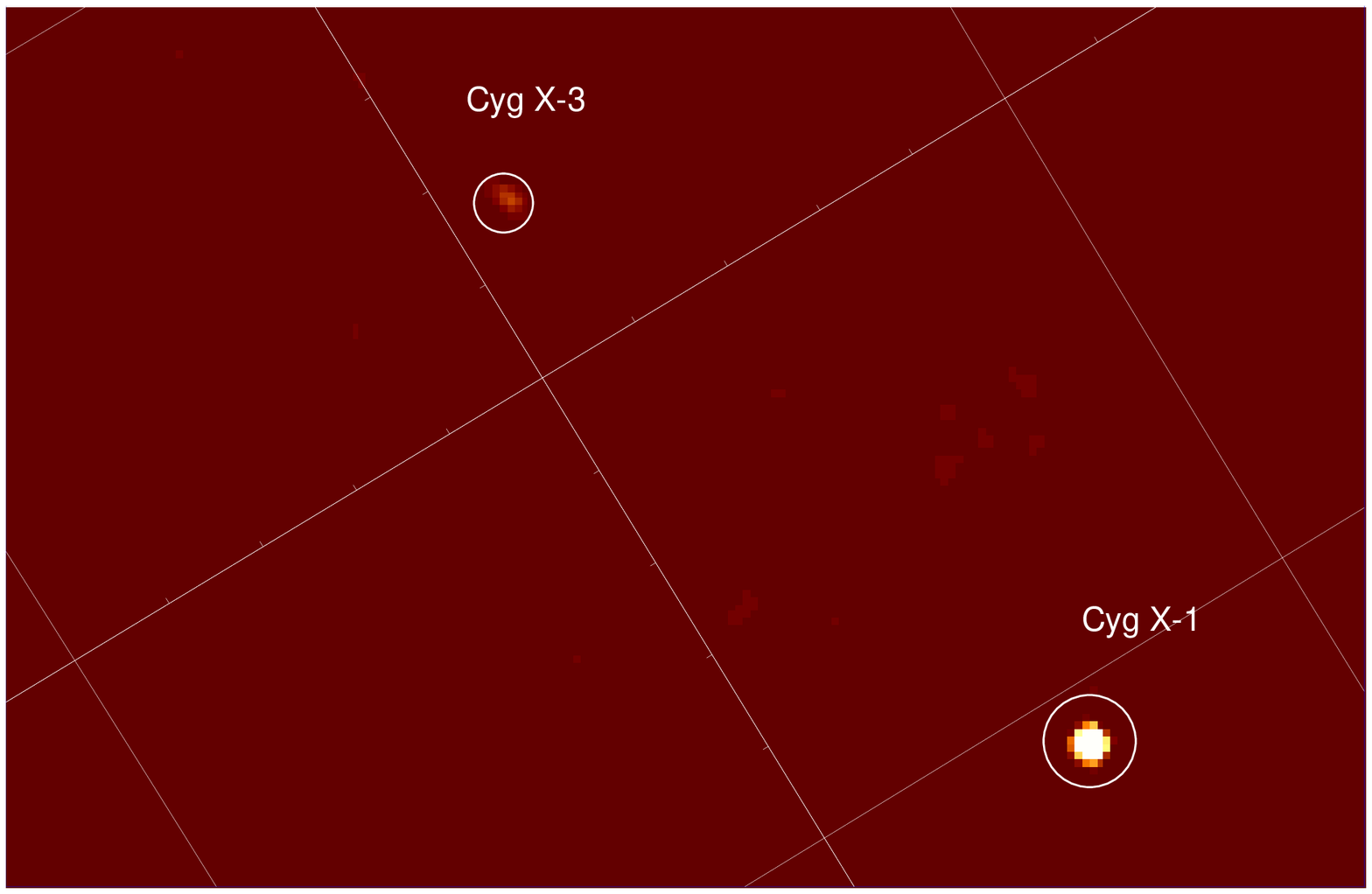}

      \caption{15-40 keV (top) and 40-100 keV deconvolution images of IBIS/ISGRI observations during revolution 16. They result from the sum of 56 Science Window of duration varying between 1800 and 3600 seconds for a total of $\sim$ 147 ksec. The confidence levels start from 10 $\sigma$ and are linearly spaced. The white grid represents Galactic Coordinates, the lines are spaced by 6 degrees. Cyg X-1 is at the
bottom left and is detected at $\sim$ 1000 $\sigma$ and $\sim$ 850 $\sigma$ level respectively, Cyg X-3 is on top right at $\sim$ 160 $\sigma$ and $\sim$ 55 $\sigma$ level respectively}
         \label{}
   \end{figure}


\noindent During the first weeks of observations, the instrument parameters
were frequently modified to optimize the instrument performances. Moreover,
the pointing direction frequently changed. For our analysis, we limit
ourselves to observations performed in the period between 27 November
2002 and 8 December 2002 when the satellite was performing staring
observations on Cyg X-1. Results of later observations on December
22-23 are described in Vilhu et al. (2003).

\noindent During the selected period the total observing time is $\sim$
$5.6 \times 10^5$ sec
but due to the off axis position, the effective exposure time on Cyg X-3
is $\sim 3.2 \times 10^5$ sec. These observations were performed during
4 orbits (revolutions) and they were split into separate 'science windows'
lasting 1800 to 3600 seconds (see Table 1 for details). The distance of
Cyg X-3 from the pointing direction was about 8.8$^{\circ}$. This distance
corresponds to a coded area or relative sensitivity of 57 $\%$ for ISGRI.

\noindent The images produced by the telescope were analyzed using the ISDC
public software and software developed at our institute for calibration purposes.
The images were deconvolved using the procedures described in Goldwurm et al.
(2003) to recover source position and flux.

  \begin{table*}
      \caption[]{Observation log and average fluxes of Cyg X-3 in the period 27/11/2002-
08/12/2002. The pointing direction was constant at $19$h $58$m $21.7$s +$35^{\circ}$ $12'$ $05''$, Cyg X-3 was 8.8$^{\circ}$ off axis with a coded fraction of
57 $\%$. The 4 periods correspond to revolution 15 to 18 and
are divided into single observations of duration varying between 1800 
and 3600 sec. Between each period there is a perigee passage lasting about
half a day.
The shorter exposure time in revolution 17 is due to the absence
of reliable data during a long period (see also Figure 2). We estimated the average fluxes
in milliCrab using Crab nebula observations at a similar off axis angle and added a 5 $\% $ systematic error.}
         \label{}
     $$ 
         \begin{array}{ccccc}
            \hline
            \noalign{\smallskip}
             {\rm Obs. Start-Stop} & {\rm Duration}  & {\rm Eff. exp.} &  15-40 {\rm ~keV~Flux}  & 40-100 {\rm ~keV~Flux}\\
            \noalign{\smallskip}
            {\rm (JD-2440000)} & {\rm (sec)}  &  {\rm (sec)} & {\rm (mCrab)} & {\rm (mCrab)}\\
            \noalign{\smallskip}
            \hline
            \noalign{\smallskip}
      52605.9-52608.5 &  153340 & 87400 & 125 \pm 6 & 60 \pm 3 \\
      52608.9-52611.5 & 146810  &  83680 & 137 \pm 3.5& 58 \pm 3 \\
      52611.9-52612.3 & 81230 & 46300 & 150 \pm 7.5 & 70 \pm 3.5   \\
      52614.9-52617.5 & 179040 &  102050 & 137 \pm 3.5 & 59 \pm 3   \\
            \noalign{\smallskip}
            \hline
         \end{array}
    $$ 
\end{table*}

   \begin{figure*}
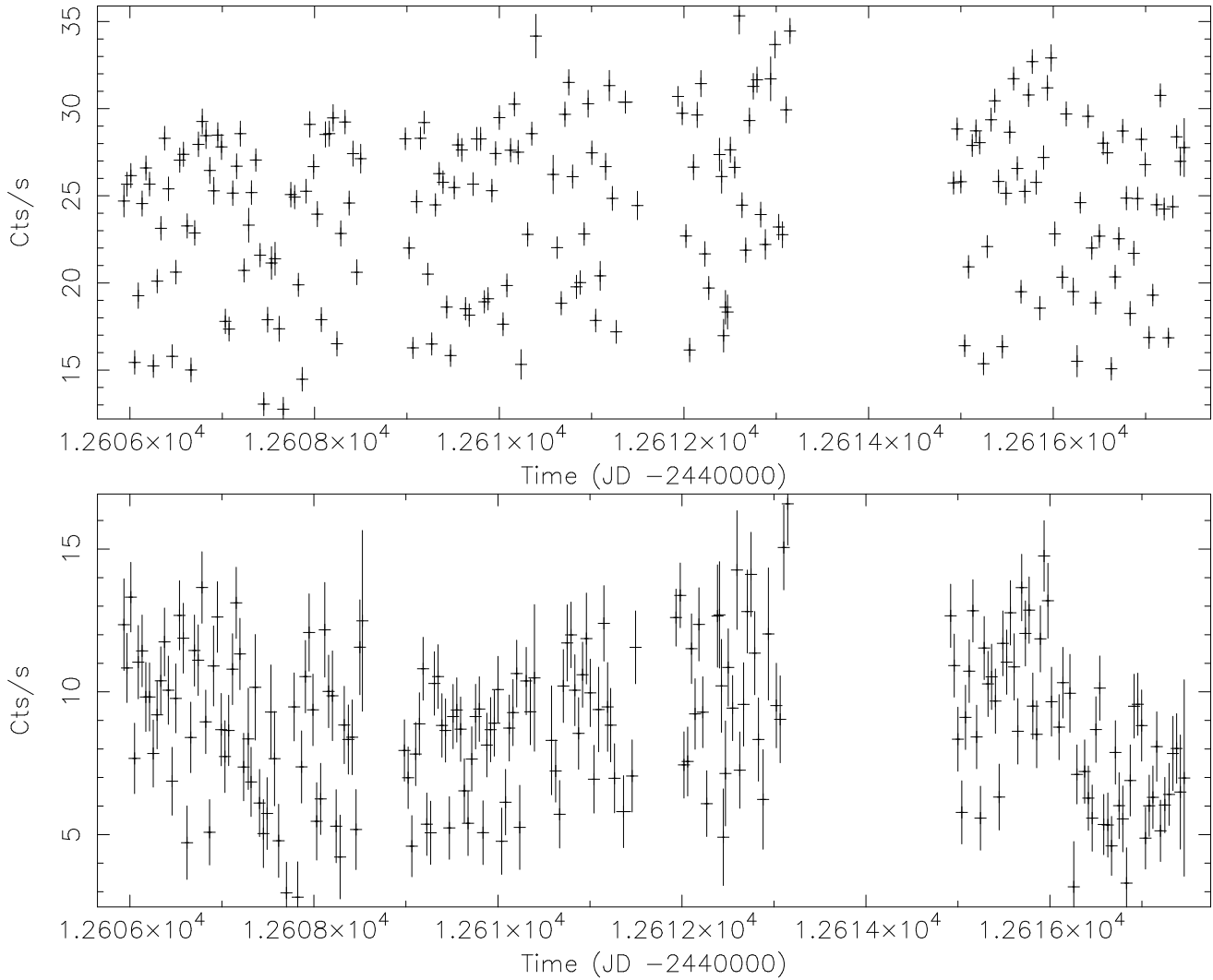

   \centering
   \includegraphics[width=7cm,angle=270]{INTEGRAL22_f2_1.ps}
   \includegraphics[width=7cm,angle=270]{INTEGRAL22_f2_2.ps}
   \caption{Cyg X-3 15-40 keV (top panel) and 40-100 keV (bottom panel) light curves for
revolutions 15 to 18. The horizontal time scale is in days. Each point represents a science window of duration varying between 1800 and 3600 seconds.Periodic gaps in the data are due to radiation belt passes while
the longer gap at day 1.2614 $\times$ 10$^4$ is due to a data loss. The point to point variation reflect the 4.8-hour modulation.}
              \label{FigGam}%
    \end{figure*}

\section {Imaging Analysis and Light Curves}

\noindent For every science window we produced deconvolved images in three
energy bands: 15-40 keV, 40-100 keV and 100-200 keV. The images of the same
revolution were then summed to attain higher sensitivity (see Fig. 1). In
the two lower energy bands Cyg X-3 is detected at a level of confidence
between 10 and 20 $\sigma$ in the majority of the science windows.
The offset of the measured position with respect to the position in the
catalog is between 0 and 4 arcmin (i.e. less than a pixel), which is
compatible with expected results for an off axis source of this intensity. The main count rate in the 15-40 keV band varied
between 35 cts/s and 10 cts/s, while the 40-100 keV count rate was between
15 cts/s and 2 cts/s. Correlated variations in the 2 energy bands on the
time-scales of days are clearly visible. In the 100-200 keV energy band
Cyg X-3 was not consistently detected in single exposure while it was
detected in the summed images lasting 150 ksec or more at a level of
$\sim$ 10$\sigma$. However the flux value is difficult to estimate as it is
near the background level. We therefore limited our analysis to the first
two energy bands. The flux ratio between the two lower energy bands indicate
a very soft spectrum with photon index $\alpha \sim$ 3.

\noindent  We estimated Cyg X-3 flux using the ISGRI count rate of the
Crab nebula at an off axis angle similar to the Cyg X-3 one in our
observations. From these first results it is apparent that the source was in
the high hard X-ray flux state seen by BATSE, its flux being stronger
than 100 mCrab in the low energy band.

\subsection{Period folding}

\noindent The evolution of the average source flux is shown in Figure 2.
Variations around the average value on the time-scale of less than a day are
clearly visible in the light curve at both energies. The point-to-point
variations reflect the effect of the 4.8-hour modulation. We folded both
source light curves with a 4.8-hour period (Figure 3). The shape of the
folded light curves are found to be very similar to the long term average
shape of the light curve obtained at lower energy and are fitted nicely
with the normalized X-ray template of van der Klis $\&$ Bonnet-Bidaud (1989).
Small variations are however visible in the rising part of the curve.
Therefore the arrival time of the minimum of the folded light curve was
derived by cross-correlating the folded data with the standard template in
the restricted phase range of [0.75 --1.25].  To take into account systematic
uncertainties, the statistical errors were scaled by a constant factor to
make them fit the template with a reduced $\chi^{2}$ of 1.0. This analysis
yielded a heliocentric minimum arrival time of (HJD 2452606.06446 $\pm$
0.0017) in the 15-40 keV range. Fitting with the overall light curve does not
introduce a significant shift. The new INTEGRAL arrival point is plotted in
Figure 4 alongside with previous results (see Singh et al. 2002) and
fits nicely with published ephemeris of Cyg X-3 thus validating our analysis.
A new ephemeris was computed by fitting a cubic function indicated by
previous studies (van der Klis $\&$ Bonnet-Bidaud 1989, Singh et al. 2002).
The INTEGRAL observations confirm the existence of a significant rate of
change of the period derivative. We obtain a significant nonzero value
of {\"{P}=$-(2.2\pm1.5)\times10^{-11}$yr$^{-1}$, strenghtening the
last published value of {\"P}=$-(1.3\pm1.6)\times10^{-11}$yr$^{-1}$ (Singh
et al. 2002). Future INTEGRAL monitoring during the core program will allow
to gain better precision on this term.

  \begin{figure}
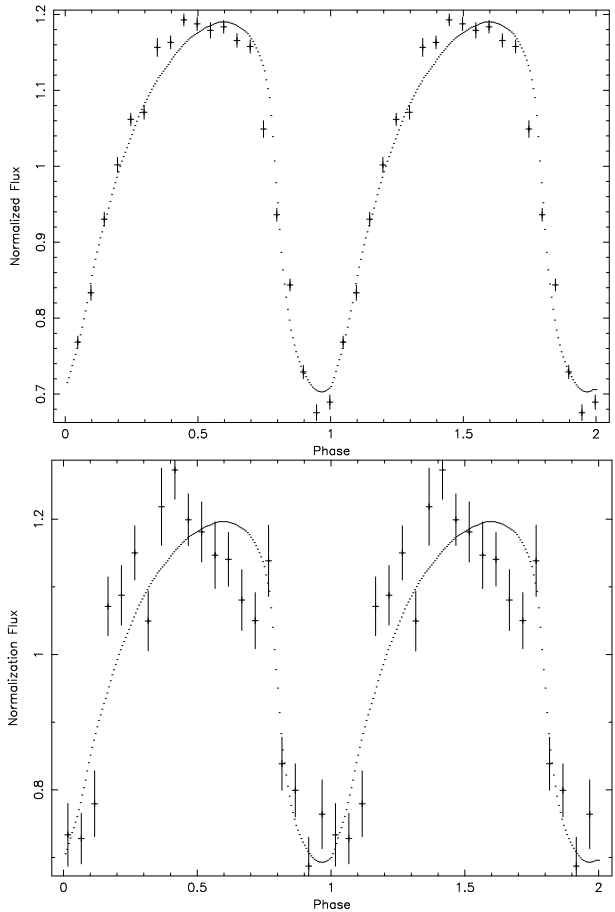

   \centering
   \includegraphics[width=6 cm, angle=270]{INTEGRAL22_f3_1.ps}
 \includegraphics[width=6 cm, angle=270]{INTEGRAL22_f3_2.ps}
      \caption{Cyg X-3 light curves in the 15-40 keV (top panel) and 40-100 keV (bottom
panel) energy bands folded to the 4.8-hour period using the template from van der Klis
\& Bonnet-Bidaud (1989).}
         \label{}
   \end{figure}

\noindent We checked for the possible presence of a time delay between 
the hard (40-100 keV) and the soft energy (15-40 keV) light curves. Indeed,
Matz (1997) reported a 6-7 $\sigma$ evidence of a constant time delay of
$\sim$ 20 min between XTE/ASM (2-12 keV) and OSSE (44-130 keV) data.
We correlated the two light curves with the template in the restricted
phase range of [0.75-1.25] and measured the difference in the minimun arrival times
obtained. We found no measurable
time delay between the two light curves. The (hard-soft)
difference in the minimum arrival times is estimated to be $\Delta$t=
$-$(5.5$\pm$8.6) minutes where the error bar is at a 3$\sigma$ level. We
therefore do not confirm the presence of a time delay between soft and hard
X-rays at this level.

   \begin{figure}
   \centering
   \includegraphics[width=6.8cm,height=8.5cm,angle=270]{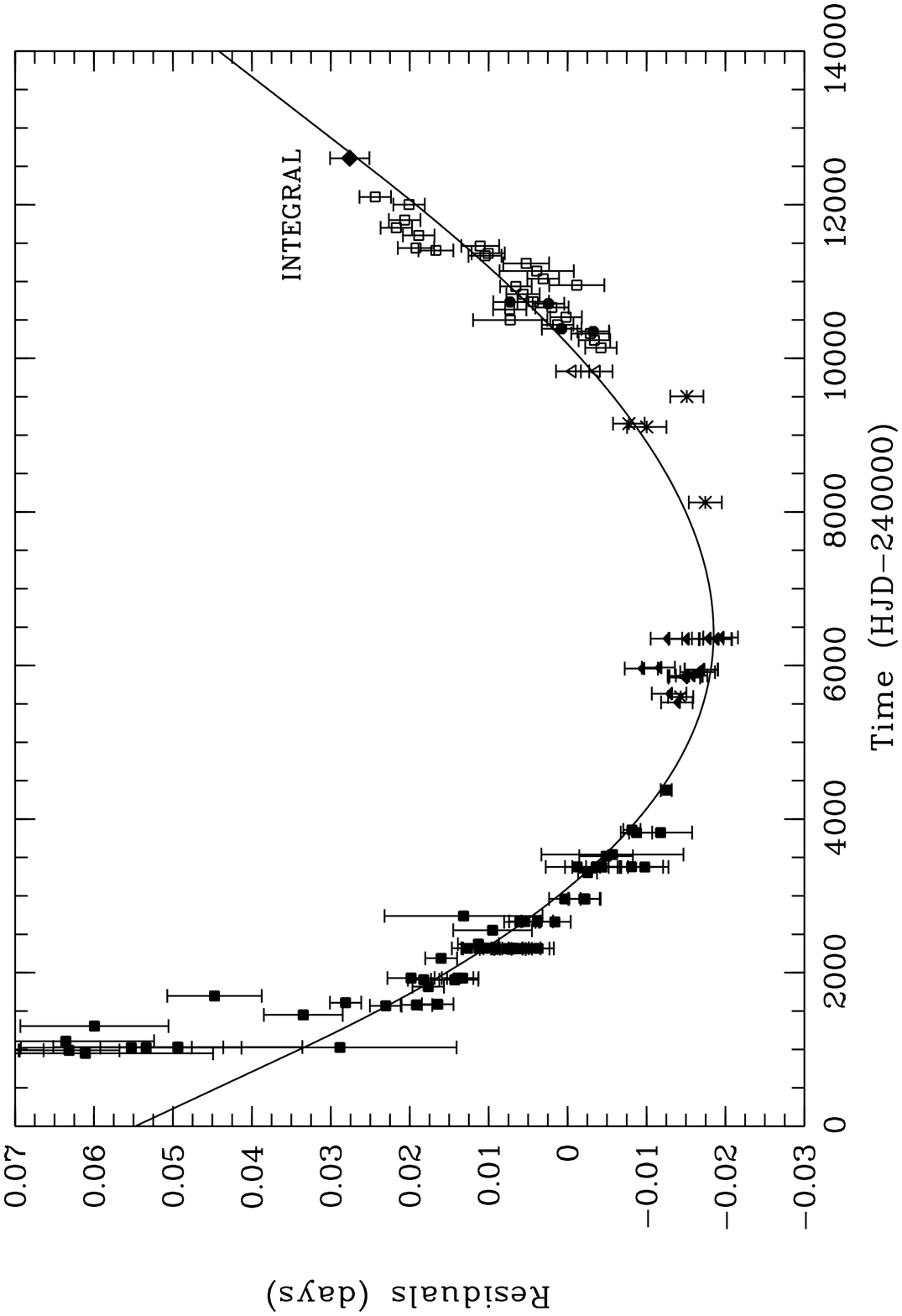}
   \caption{Cyg X-3 arrival times residuals with respect to the linear 
   ephemeris of Singh et al. (2002). Different symbols are for data earlier than 
   1983 (filled squares), EXOSAT (filled triangles), Tenma, GINGA and ASCA 
   (asterisks), ROSAT (open triangles), XTE/ASM (open squares), BeppoSAX (filled circles)
   and INTEGRAL (filled lozenge). Also shown is the best fitted cubic ephemeris 
   with a significant slowing down of the period derivative.}
              \label{FigGam}%
    \end{figure}

\section{Conclusions}

\noindent We report on first results of the ISGRI/IBIS detector
onboard the INTEGRAL observatory for the peculiar microquasar
Cyg X-3. The source was serendipitously observed during Cyg X-1
Performance and Verification Phase observations and during the observations
we selected it was off axis by 8.8$^{\circ}$.

\noindent We demonstrate that, despite the {\sl a priori} unfavorable
position of the source in the FOV, we are able to obtain high quality
results. Even if at this stage of data analysis, the results are still
preliminary, we demonstrate that the 4.8-hour modulation is clearly visible.
Moreover the absolute time of the phase zero is fully consistent
with previously published data. No measurable time delay could be found
between the two light curves.

\noindent The light curve shape that we obtain for the two energy bands
possibly indicates that the same physical mechanism is responsible for
the bulk of photon emission between 15 and 100 keV.

\noindent Even at this early stage of the mission, we demonstrate
the IBIS/ISGRI detector capability of producing scientific results for
relatively bright ($>$50 mCrab), off axis targets. This capability is
one of the most important IBIS/ISGRI characteristics and its exploitation
will be very important throughout the mission.

\end{document}